\documentclass{nature}
\usepackage{graphicx}
\usepackage{color}
\usepackage[colorlinks=false, pdfborder={0 0 0}]{hyperref}
\usepackage{rotating}

\def\kpc{{\rm\,kpc}}
\def\Mpc{{\rm\,Mpc}}
\def\msun{{\rm\,M_\odot}}
\def\kms{{\rm\,km\,s^{-1}}}
\def\ltsima{$\; \buildrel < \over \sim \;$}
\def\simlt{\lower.5ex\hbox{\ltsima}}
\def\gtsima{$\; \buildrel > \over \sim \;$}
\def\simgt{\lower.5ex\hbox{\gtsima}}

\newcommand*{\ditto}{---''---}

\title{Velocity anti-correlation of diametrically opposed galaxy satellites in the low $z$ universe}

\author{Neil G. Ibata$^{1}$, Rodrigo A. Ibata$^{2}$, Benoit Famaey$^{2}$, Geraint F. Lewis$^{3}$}

\begin{document}

\maketitle

\begin{affiliations}
\item Lyc\'ee international des Pontonniers, 1 rue des Pontonniers, F-67000 Strasbourg, France.
\item Observatoire astronomique de Strasbourg, Universit\'e de Strasbourg, CNRS, UMR 7550, 11 rue de l'Universit\'e, F-67000 Strasbourg, France
\item Sydney Institute for Astronomy, School of Physics, A28, The University of Sydney, NSW 2006, Australia
\end{affiliations}

\newpage	

\begin{abstract}
Recent work has shown that both the Milky Way and the Andromeda galaxies possess the unexpected property that their dwarf satellite galaxies are aligned in thin and kinematically coherent planar structures\cite{2005A&A...431..517K,2012MNRAS.423.1109P,2013MNRAS.435.1928P,2013MNRAS.435.2116P,2013Natur.493...62I,2013ApJ...766..120C,2014ApJ...784L...6I}. It is now important to evaluate the incidence of such planar structures in the larger galactic population, since the Local Group may not be a sufficiently representative environment. Here we report that the measurement of the velocity of pairs of diametrically opposed galaxy satellites provides a means to determine statistically the prevalence of kinematically coherent planar alignments. In the local universe (redshift $z<0.05$), we find that such satellite pairs out to a galactocentric distance of $150\kpc$ are preferentially anti-correlated in their velocities (99.994\% confidence level), and that the distribution of galaxies in the larger scale environment (beyond $150\kpc$ and up to $\approx 2 \Mpc$) is strongly elongated along the axis joining the inner satellite pair ($>7\sigma$ confidence). Our finding may indicate that co-rotating planes of satellites, similar to that seen around the Andromeda galaxy, are ubiquitous in nature, while their coherent motion also suggests that they are a significant repository of angular momentum on $\sim 100\kpc$ scales.
\end{abstract}

The satellite galaxies of the Milky Way have long been known to be preferentially located close to a plane\cite{1976MNRAS.174..695L}, but this observation could be dismissed as a mere coincidence. However, as faint galaxies were uncovered in the Sloan Digital Sky Survey (SDSS)\cite{AdelmanMcCarthy:2006iz}, it became clear that our Galaxy hosts a planar structure of satellites with a close to polar orientation\cite{2007MNRAS.374.1125M,2008ApJ...680..287M,2009MNRAS.394.2223M}. The complications due to spatial incompleteness of satellite samples that complicate analyses in the Milky Way are largely alleviated when observing the next nearest giant galaxy, Andromeda (M31). The presence of a vast plane of co-rotating dwarf galaxies was recently detected in that galaxy thanks to new photometric\cite{2009Natur.461...66M,2012ApJ...758...11C,2014ApJ...780..128I} and spectroscopic\cite{2012ApJ...752...45T,2013ApJ...768..172C} surveys of its halo. A full 50\% of the dwarf galaxies around M31 belong to this structure\cite{2013Natur.493...62I,2013ApJ...766..120C}. The satellites in the plane extend to $\sim 300\kpc$, yet they display very small scatter in the direction perpendicular to the plane ($12.6\kpc$), and they also possess coherent kinematics, suggestive of common rotation about their host. Recent analyses have also uncovered possible galaxy alignments in the M81 system\cite{2013AJ....146..126C}, and in the NGC~3109 association\cite{2013A&A...559L..11B}. Such satellite alignments may arise naturally if dwarf galaxies formed out of tidal debris left over by ancient galaxy mergers\cite{2012MNRAS.423.1109P,2013MNRAS.431.3543H}, but this scenario remains difficult to reconcile with the high dark matter content deduced for these objects\cite{2013pss5.book.1039W}. Although the reality of the planar structures in the Local Group is now firmly established, they may represent a fossil of the particular dynamical formation history of the Milky Way and Andromeda systems\cite{2013MNRAS.436.2096S}. It is therefore necessary to investigate more distant systems to ascertain the true significance of these local detections. 

We devised a test (Methods) to quantify the incidence of planar systems of satellites. Beyond a few Mpc, reliable and accurate relative distance measurements are inaccessible; this means that one has to deal with two-dimensional projections of galactic systems, possessing only the radial component of velocity. We take the M31 system as a template for the search of satellite alignments, since its global structure and dynamics is currently best understood. Half of that system shows coherent rotation, which means that for orientations that are not exactly face-on, satellites on either side of the galaxy as seen on the sky will in general have opposite velocities relative to the host (i.e. the velocities of the satellites will be anti-correlated). This motivates the following simple detection method: for each satellite around a given host, we check whether it possesses a counterpart that is located on the opposite side of the galaxy to within a certain tolerance angle $\alpha$ (sketch in Fig. 1a), and if it does, we determine whether the pair has correlated or anti-correlated velocities. Clearly, with circular orbits, no contamination and perfect data, all pairs will be anti-correlated if they all belong to co-rotating planes. 

As a control, we first applied this test to the large Millennium II simulation (MS2) of structure formation and evolution\cite{2009MNRAS.398.1150B,2013MNRAS.428.1351G}, which reflects our best theories of galaxy formation in $\Lambda$ Cold Dark Matter ($\Lambda$CDM) cosmology\cite{2011ApJS..192...18K}. We find that diametrically opposite pairs of bright satellites selected from that simulation (Methods) show no kinematic coherence, with roughly equal numbers of correlated and anti-correlated pairs for all $\alpha$ (filled circles in Fig. 1b). Note however that with a very different satellite selection strategy, a slight preference for co-rotating satellites can be found\cite{2014MNRAS.438.2916B}. To analyse the behaviour of our statistic in the presence of a contaminating background, we forced different fractions of satellites around M31-like hosts in MS2 (Methods) to lie within a randomly-chosen rotating plane. Fig. 1c shows the anti-correlation of the satellite pairs as a function of the dominance of planar configurations; evidently a measure in real galaxies of the fraction of anti-correlated pairs has the potential to reveal whether planar satellite alignments are common. 

We therefore applied this test to the SDSS, which currently gives the most complete view of the nearby universe. Since we wish to investigate the environment around galaxies similar to the Milky Way and Andromeda, we select from the ``NYU Value-Added Galaxy Catalog''\cite{2005AJ....129.2562B} isolated host galaxies (no brighter neighbour within a distance of $500\kpc$ and within a velocity difference of $1500\kms$) with magnitudes in the range $-23\le {\rm M_r} \le -20$. Cosmological parameters from the Planck mission are assumed\cite{2013arXiv1303.5076P}. To ensure a clean sample of satellites, we select hosts up to a redshift $z=0.05$ (beyond this, few faint satellites are detected), remove hosts closer than $z=0.002$ to avoid noisy measurements, and remove all galaxies with velocity uncertainties greater than $25\kms$. The satellites themselves are any galaxy one magnitude or more fainter than the host, but brighter than ${\rm M_r}=-16$, within the radial range $20\kpc<R<150\kpc$ (again to be similar to the M31 analysis\cite{2013Natur.493...62I}). We further require that the satellites at projected distance $R$ lie within a velocity of $300 \exp[-(300\kpc/R)^{0.8}]\kms$ (see Methods, Extended Data).

As for the MS2 analysis, we retain only those satellites whose direction of motion with respect to their hosts is well-resolved; since we imposed an upper velocity uncertainty of $25\kms$ for both hosts and satellites, we require a minimum velocity difference of $\sqrt{2} \times 25\kms$. There are $380$ galaxy systems in the SDSS that pass these requirements.

Various choices for $\alpha$ are examined in Fig. 2. As our toy model shows (Methods, Fig. 1b), the highest contrast between anti-correlated and correlated satellite pairs should be found for small $\alpha$. There is an inevitable trade-off between the number of satellite pairs that pass the selection criteria and the contamination fraction suffered by the sample. Strict selection gives good contrast, but poor statistics; lenient selection gives poor contrast, but good statistics. Optimal significance will therefore lie at an intermediate tolerance angle, but given the unknown density and kinematic properties of both the normal and putative disk-like satellite populations, we believe the best strategy is to allow the data themselves to guide our choice of $\alpha$. Fig. 2a shows that with $\alpha=8^\circ$, 20 out of 22 pairs are anti-correlated, implying 99.994\% ($>4\sigma$) significance; these systems are listed in Table 1. (With a less strict velocity cutoff of $\sqrt{2}\times20\kms$, 21 out of 23 pairs are anti-correlated). High significance is found out to $\alpha=15^\circ$, although as expected in the presence of non-planar ``contaminants'', the significance decreases with increasing $\alpha$. By comparison to the simulations where a disk population was added to MS2 (Fig. 1c), the observed ratio of anti-correlated to correlated pairs ($>2.7$ at 99\% confidence) at $\alpha=8^\circ$ suggests that a fraction of $> 60$\% of satellites reside in planes, although we stress that this constraint is weak as this disk model is simplistic. Thus we have found that the average giant galaxy in the SDSS is consistent with our M31 template. While the SDSS spectroscopic observing strategy produced certain spatial biasses\cite{2002AJ....124.1810S}, it seems extremely improbable that such biasses could artificially cause an over-abundance of anti-correlated satellite pairs as found here. 

Fig. 3 highlights a possible correlation between the direction defined by the satellite alignment and the large-scale structure surrounding the hosts. An elongated overdensity of galaxies appears aligned along the axis of the satellite pair, extending out to $\approx 10$ times the distance of the selected pair (Figs.~3a,b). This is consistent with what we see in the Local Group, where the M31 satellite alignment points within $1^\circ$ of the Milky Way. Although these filaments are much thicker than the planes around galaxies, it is possible that this reveals the influence of large scale structures on the dynamics of the smaller satellite system. Furthermore, in MS2 the larger-scale environment around anti-correlated pairs shows no strong preferential direction, and neither does the environment around SDSS correlated pairs (Fig. 3c). While it remains possible that the large-scale elongation of the galaxy distribution along the direction of the galaxy pairs is an artefact of the SDSS target selection, this seems unlikely given the random orientation of the satellite pairs on the sky.

Table 1 also lists the angular momentum of the satellite pairs, calculated using the projected distance, line of sight velocity and estimated stellar masses\cite{2005AJ....129.2562B} of these galaxies. For comparison, the total angular momentum of our Galaxy in stars is $|L_\star| \sim 9\times 10^{13}\msun\kms\kpc$ (approximating it as an exponential disk\cite{2011MNRAS.414.2446M}). Thus the angular momentum contained in the stellar component of the aligned satellites we have identified (mean of the $\alpha=8^\circ$ sample: $\langle A |L_\star| \rangle=8.3\times10^{13}\msun\kms\kpc$, where $A$ is a sign flag explained in Table~1) is comparable to the angular momentum in a giant galaxy's stellar disk. This suggests that these coherent structures make a significant contribution to the angular momentum budget on galaxy halo scales ($\sim 100\kpc$), although a better understanding of their incidence and physical properties is required to quantify their importance.

Our tests were constructed using MS2 as a control sample to predict what should have been a priori expected in $\Lambda$CDM cosmology. Just as this paradigm did not predict the planes observed in the Local Group\cite{2014ApJ...784L...6I}, it did not a priori predict the velocity correlations presented here. It should be noted, however, that MS2 contains only dark matter, and future large cosmological simulations that include detailed baryonic physics should be performed to see if the discrepancies can be alleviated. While it is as yet uncertain whether the pairs of satellites detected here actually form part of kinematically coherent planes, their velocity anti-correlation, alignment with larger scale structures, and high angular momentum are all unexpected properties of the Universe that will require explanation.

\section*{Methods Summary}

Our test uses satellites that are diametrically opposite each other around their host to quantify the incidence of rotating planar alignments. The signature of coherent rotation is an enhancement in the number of anti-correlated satellites. Using a small tolerance angle (Fig. 1a) and a minimum velocity difference, samples can be generated with a higher probability of containing edge-on planar structures, if they are present. We first test this method on a simple toy model, to show the expected behaviour to choices of the tolerance angle parameter $\alpha$ (Fig. 1b): the contrast of the planar component is seen to decrease with increasing $\alpha$, suggesting that small values of $\alpha$ should preferably be used for the tests. To construct a somewhat more realistic model, we select galaxies and their satellites from the Millennium II cosmological simulation, and reassign some of the satellites to planar structures. The selection process for hosts and satellites is kept as close as possible to the selections applied to the observed SDSS sample.

\newpage

\newcommand{\apj}{Astrophysical Journal}
\newcommand{\apjs}{Astrophysical Journal Supplement}
\newcommand{\apjl}{Astrophysical Journal Letters}
\newcommand{\aj}{Astronomical Journal}
\newcommand{\mnras}{Monthly Notices of the Royal Astronomical Society}
\newcommand{\nat}{Nature}
\newcommand{\pasa}{Publications of the Astronomical Society of Australia}

\newpage

\begin{addendum}
 \item
Funding for the SDSS and SDSS-II has been provided by the Alfred P. Sloan Foundation, the Participating Institutions, the National Science Foundation, the U.S. Department of Energy, the National Aeronautics and Space Administration, the Japanese Monbukagakusho, the Max Planck Society, and the Higher Education Funding Council for England. The SDSS Web Site is http://www.sdss.org/. The Millennium-II Simulation databases used in this paper and the web application providing online access to them were constructed as part of the activities of the German Astrophysical Virtual Observatory (GAVO).
\item[Author contributions] All authors assisted in the development and writing of the paper. N.I. primarily contributed to the development of the test for planar alignments, while R.I. implemented this test on the SDSS galaxy catalogue.  
\item[Competing Interests] The authors  have no competing financial interests.
\item[Correspondence] Reprints and permissions information is available at www.nature.com/reprints. Correspondence and requests for materials should be addressed to N.I. (neil.ibata@gmail.com).
\end{addendum}

\newpage

\noindent
{\bf Table 1:} The redshift $z$, positions (right ascension and declination in degrees), absolute magnitudes ${\rm M_r}$ and radial velocities $v$ (in $\kms$) of the hosts (superscript ``h'') and the satellites (superscript ``S1'' or ``S2''), for the sample selected with a tolerance angle of $\alpha=8^\circ$. The final column lists the sum of the angular momentum of the stellar component of both satellites $|L_{\star}| = |L_{\star}^{\rm S1} + L_{\star}^{\rm S2}|$, multiplied by a sign flag $A$, where $A=1$ implies that the pair have anti-correlated velocity and $A=-1$ that the velocity is correlated.

{\renewcommand{\arraystretch}{0.5}
\begin{sidewaystable}
{\small
\begin{center}
\caption{The SDSS systems with diametrically opposite satellite pairs for $\alpha=8^\circ$. }
\begin{tabular}{rrrrrrrrrrrrr}
\hline
\hline
z     & 
       RA$^{\rm h}$&
                   dec$^{\rm h}$&
                               ${\rm M_r^h}$&
                                       RA$^{\rm S1}$&
                                                   dec$^{\rm S1}$&
                                                               ${\rm M_r^{S1}}$&
                                                                       $v^{S1}$&
                                                                                RA$^{\rm S2}$&
                                                                                            dec$^{\rm S2}$&
                                                                                                        ${\rm M_r^{S2}}$&
                                                                                                                $v^{S2}$&$A |L_{\star}| / 10^{13}$\\
      & deg& deg&  mag& deg& deg&  mag&km$/$s&deg&deg&  mag&km$/$s&$\msun\kms\kpc$\\
\hline
0.0324& 318.182771&  -0.387524& -20.01& 318.172302&  -0.395010& -17.46&   43.& 318.218964&  -0.366205& -18.31& -131.&  1.2\\
0.0395&   3.140439&  -0.048469& -22.00&   3.102046&  -0.071687& -18.63& -113.&   3.166931&  -0.036225& -19.74&   64.&  5.9\\
\ditto&   \ditto  &  \ditto   & \ditto&   \ditto  &   \ditto  & \ditto&\ditto&   3.182354&  -0.028088& -19.35&   66.&  9.7\\
0.0197& 181.112776&   1.895961& -22.84& 181.153046&   1.892648& -17.90&  189.& 181.054535&   1.900154& -19.29& -159.& 12.1\\
0.0476& 129.700626&   4.126124& -22.02& 129.688538&   4.113726& -20.12& -182.& 129.715744&   4.139352& -19.39&  161.&  8.3\\
0.0443& 230.341251&   5.066835& -22.23& 230.337646&   5.045036& -18.92&  107.& 230.342621&   5.082857& -19.08& -116.&  3.8\\
0.0227& 203.116383&   7.316453& -22.46& 203.134415&   7.294215& -20.49&  -52.& 203.106171&   7.331840& -18.68&  195.&  7.3\\
0.0197& 210.702609&   9.341378& -22.08& 210.619705&   9.315668& -17.20& -180.& 210.779785&   9.363044& -20.82&   86.& 31.4\\
0.0360& 244.995300&  10.493549& -22.41& 244.980530&  10.486047& -18.54&  -57.& 245.008408&  10.499287& -21.12&  231.& 30.8\\
0.0266& 149.548260&  15.946962& -21.78& 149.566422&  15.919011& -18.66&  211.& 149.534317&  15.963491& -20.39&  -96.&  9.8\\
0.0390& 241.663921&  17.761186& -22.50& 241.681305&  17.727509& -18.62& -180.& 241.651321&  17.796453& -20.56&   40.& 14.7\\
0.0208& 176.008968&  19.949823& -22.68& 176.050766&  19.942766& -17.97&  -41.& 175.986832&  19.955805& -18.79&   62.&  1.4\\
0.0371& 156.234227&  20.402742& -22.05& 156.282028&  20.396730& -18.59&   53.& 156.221756&  20.403419& -19.77& -125.&  5.8\\
0.0170& 157.320107&  26.099230& -21.48& 157.346039&  26.070440& -18.94& -107.& 157.256622&  26.153496& -18.20&   87.&  2.8\\
0.0272& 168.791680&  31.033707& -22.27& 168.785461&  31.002020& -20.08&  212.& 168.804108&  31.079819& -17.95& -166.& 23.8\\
0.0428& 170.378180&  33.957669& -22.57& 170.338287&  33.955376& -19.55& -111.& 170.402756&  33.958282& -18.79&  -98.& -6.6\\
0.0277& 167.171535&  36.161068& -21.50& 167.216934&  36.116596& -18.07&  -90.& 167.115128&  36.206646& -19.52&   94.&  7.8\\
0.0316& 247.216940&  42.812009& -21.74& 247.204544&  42.805214& -19.21& -116.& 247.243317&  42.828125& -20.59&   45.&  7.0\\
0.0369& 164.010031&  44.395565& -21.29& 163.993439&  44.380817& -18.96&  111.& 164.039291&  44.428234& -19.13&  159.& -8.3\\
0.0375& 128.717100&  44.637116& -21.91& 128.674500&  44.593529& -19.51&  134.& 128.751175&  44.671597& -19.02& -156.&  9.9\\
0.0245& 195.793476&  47.393865& -21.55& 195.792770&  47.381126& -19.08&  -97.& 195.788147&  47.447929& -18.75&   51.&  2.1\\
0.0233& 143.529090&  67.431139& -21.35& 143.660706&  67.374802& -18.20&  -90.& 143.446182&  67.477814& -18.46&   52.&  2.1\\

\hline
\hline
\end{tabular}
\end{center}
}
\end{sidewaystable}
}

\newpage

\begin{figure}
\includegraphics[angle=0, bb=70 315 550 483, clip,  width=\hsize]{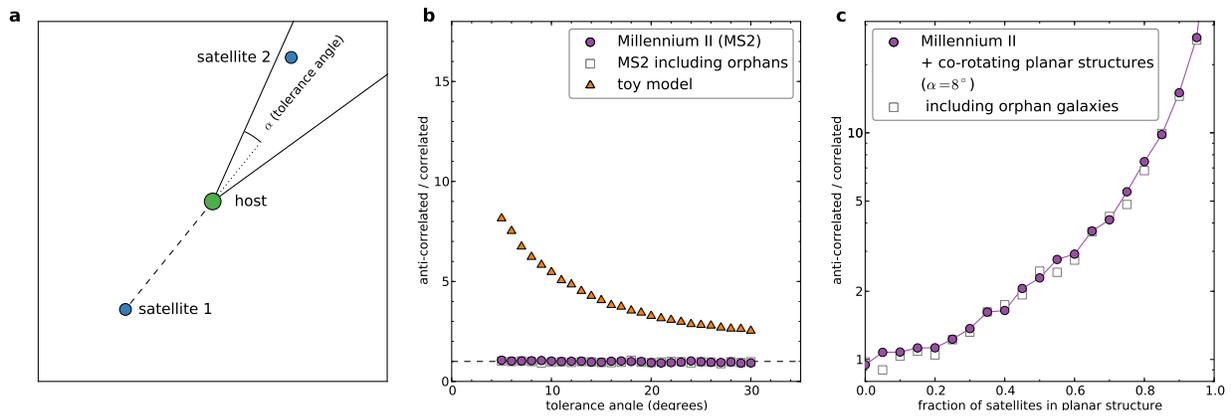}
\caption{{\bf Satellite correlation test.} 
{\bf a}, Sketch of the satellite selection process. 
{\bf b}, The fraction of anti-correlated to correlated satellite pairs in MS2 (rejecting or including ``orphan'' galaxies) is consistently very close to 1, independent of $\alpha$. However, the simple toy model (Methods) shows a decline of the ratio with increasing $\alpha$. 
{\bf c}, Fraction of anti-correlated galaxy pairs as a function of the fraction of satellites in the rotating planar population (using $\alpha=8^\circ$, the most significant peak in Fig. 2c). In the absence of a planar component, equal numbers of correlated and anti-correlated satellites should be detected. However, the ratio increases as expected as the planar component is made more significant.}
\end{figure}

\begin{figure}
\includegraphics[angle=0, bb=70 315 550 483, clip,  width=\hsize]{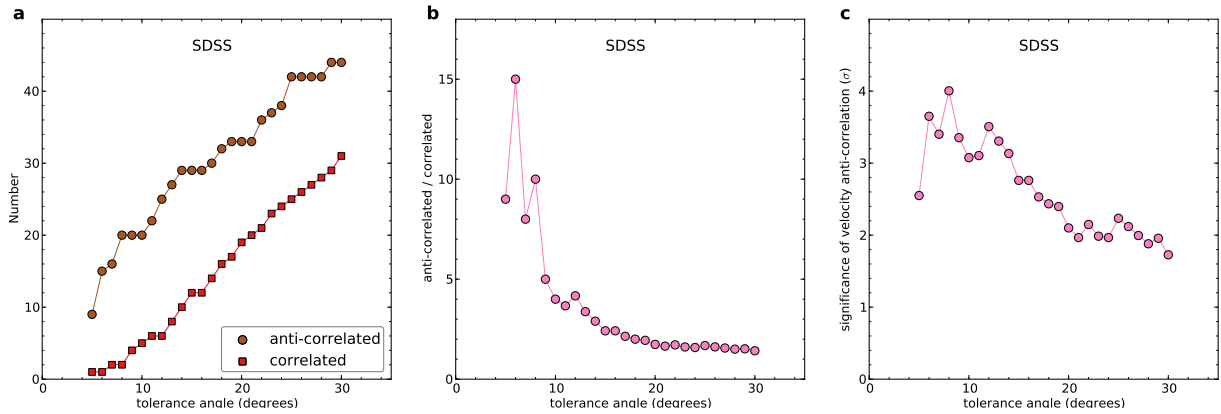}
\caption{{\bf Anti-correlated satellites in the SDSS.}
{\bf a}, The number of satellite pairs that have correlated and anti-correlated velocities is shown as function of the tolerance angle. There is a clear surplus of anti-correlated pairs for all angles considered. 
{\bf b}, This fraction shows an overall decline with increasing tolerance angle, reaching $2.4$ at $15^\circ$, which we consider the maximum useful opening angle given the low number of satellite pairs in the SDSS. 
{\bf c}, The significance (in units of standard deviation) of the excess of anti-correlated satellite pairs. The most significant peak has significance $>4\sigma$ at an  opening angle of $8^\circ$.}
\end{figure}

\begin{figure}
\includegraphics[angle=0, bb=70 315 550 483, clip,  width=\hsize]{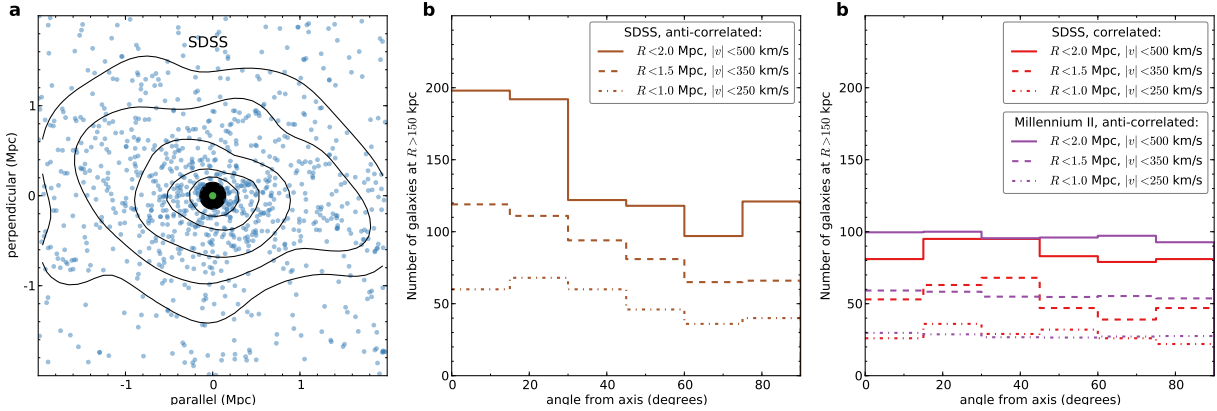}
\caption{{\bf Correlation with environment.} 
{\bf a}, Superposition of SDSS galaxies (within $500\kms$) that surround the hosts of the satellite pairs with anti-correlated velocities (using $\alpha=15^\circ$). Each field is rotated so that the receding satellite lies on the positive abscissa. A clear horizontal feature is found out to $\sim 2\Mpc$; this result remains robust for various subsamples and parameter choices. (The black disk shows a radius of $150\kpc$). 
{\bf b}, The angular distribution of the galaxies in (a), rejecting galaxies within $R<150\kpc$. The significance of the peaks for the $R < 1.0, 1.5, 2\Mpc$ samples are $3.7, 4.8, 7.1\sigma$, respectively. 
{\bf c}, Applying the same procedure to the region around SDSS correlated pairs (red line, using $\alpha=20^\circ$ to build up better statistics) shows minimal correlation, as does the environment around anti-correlated pairs in MS2 (purple).}
\end{figure}

\clearpage

\begin{center}
{\Large\bf Methods}
\end{center}

\section*{Simple kinematic test on diametrically opposite satellites}

The simple statistical test we have developed is devised to allow us to quantify the frequency of satellites belonging to disk-like structures. We use primarily the distinctive property of a rotating disk-like structure that objects on opposing sides have anti-correlated velocities. The expectation from observations of M31 is that any such structures are superposed on a ``contaminating'' population of ``normal'' satellites which appear, to first approximation, to have a spherically symmetric distribution around the host. The presence of such a contaminating population, together with the fact that most galaxies beyond the Local Group have only a small number of satellites with well-measured velocities, means that at present we can only test for the alignments in a statistical manner on a sample of hosts.

Thus the challenge is to devise a means to enhance the contrast of the putative disk over a potentially dominant spherical population. Since our viewing direction on these distant systems cannot be special, on average, the median inclination of any disk-like structures will be $60^\circ$ (if we define an edge-on configuration to have inclination $90^\circ$). This naturally suggests using a test that makes use of the resulting elongation. However, we can bias a sample towards being more edge-on by selecting those systems with satellites that have radial velocities significantly different from their host galaxy. (Clearly face-on disk-like alignments will have zero velocity difference, as viewed along our line of sight).

As sketched in Figure 1a, we consider systems consisting of a massive host galaxy harbouring at least one pair of satellites. Picking each satellite in turn, we determine whether another satellite lies on the opposite side of the host within a tolerance angle $\alpha$. If both satellites possess a velocity that is significantly different from that of their host, the pair is retained for study. Motivated by the galaxy velocity uncertainties in the SDSS, we selected this minimum velocity difference parameter to be $\Delta v_{\rm min} = \sqrt{2} \times 25\kms$ in all calculations presented here (the results are qualitatively very similar for $30\kms < \Delta v_{\rm min} < 40\kms$).

To explore how the method works, we constructed first a very simple test configuration, containing 50\% of satellites in a spherical population, and 50\% in a disk-like structure. Both structures were populated with satellites with uniform probability between $20\kpc$ and $150\kpc$. The satellites in the disk rotate at $40\kms$, independent of radius, while the spherical population has an isotropic velocity dispersion of $70\kms$. This toy model is then viewed from a random direction, and two satellites are selected at random beyond a projected radius of $20\kpc$. If the two satellites lie on opposite sides of the host to within the chosen tolerance angle, and if both satellites have a velocity difference with respect to their host of more than $\Delta v_{\rm min}$, the pair is retained, and we count whether the velocities are correlated or anti-correlated. The procedure is repeated 2000 times for each tolerance angle value.

The open triangles in Fig. 1b show the ratio of the number of anti-correlated to correlated pairs as a function of the tolerance angle $\alpha$. As this simple model shows, we expect the highest contrast to be found at small tolerance angles. The selection on diametrically opposite galaxies together with the minimum velocity criterion ensures that close to edge-on configurations are preferentially selected; for this toy model, the average inclination angle for the $\alpha=10^\circ$ sample is $80^\circ$.

\section*{Construction of artificial satellite systems from the Millennium II simulation}

In order to explore further the reliability of our method to uncover genuine planar satellite alignments, we decided to construct artificial galaxy systems that we could run and test our algorithm on. For this purpose the Millennium II simulation (in particular with the semi-analytic modelling by Guo et al.\cite{2013MNRAS.428.1351G}) provides an ideal view of the expected distribution of galaxies and their satellites in a very large ($10^6 h^{-3} \Mpc^3$) volume in a $\Lambda$CDM universe. The catalogue lists the absolute r-band magnitudes, total galaxy masses, positions and velocities that are necessary for our comparison to observations. 

To create the random views of galaxies derived from the simulation we proceed as follows. We first choose a random direction from which we will view the galaxies in the Millennium II volume. A list of candidate host galaxies is generated by selecting those objects with absolute magnitudes in the range $-23\le {\rm M_r} \le -20$ (identical to the selection on the real SDSS data presented in the main text). We examine each of the candidate hosts in turn, placing the host to be studied at $10000\kms$ (the mean velocity of the SDSS sample), and then making sure that it appears isolated in projection with no brighter neighbour within $0.5\Mpc$, with velocity differences less than $1500\kms$. 

We make a list of all the neighbouring galaxies within a projected distance of $500\kpc$ and a velocity of $1500\kms$ that are at least one magnitude fainter than the host, but that are brighter than ${\rm M_r}=-16$; we will refer to these objects as ``satellites''. (We reject so-called ``orphan'' galaxies which are systems whose parent subhalo is no longer resolved in the Millennium II simulation --- but as Figs 1b and 1c show, our results remain qualitatively identical if these objects are included). For each host we then randomly draw a vector to define the normal to the planar population, and we go through the list of satellites randomly assigning them to the planar population, according to the desired planar component fraction that we wish to test for. Clearly, when testing for a planar fraction of zero, the Millennium II positions and velocities remain unaltered. For those satellites that are assigned thus to the planar component, we keep the galactocentric distance that they had in the Millennium II simulation, but place them onto the plane with a random azimuthal angle. The space velocities of the planar satellites are devised to give circular motions in the plane of the alignment, with the total velocity chosen from the circular velocity of a universal halo model\cite{Navarro:1997if} of total mass given by the virial mass of the host. 

A Gaussian random velocity of $15\kms$ (a representative value for the SDSS velocity errors) is added to the radial velocity of the host and all its satellites. Having thus reordered the three-dimensional positions of some of the satellites, we filter the sample to keep those objects that lie at projected distances in the range $20\kpc < R < 150\kpc$, and that have velocity difference of less than $300 \exp[-(300\kpc/R)^{0.8}]\kms$ with respect to the host (Extended Data Fig. 1). The brightest two satellites within the $20\kpc < R < 150\kpc$ annulus are selected for study. If the two satellites lie on opposite sides of the host to within the chosen tolerance angle, and if both satellites have a velocity difference with respect to their host of more than $\Delta v_{\rm min}$, the pair is retained, and we count whether the velocities are correlated or anti-correlated.

The entire process is then repeated for all other candidate hosts. We rerun the entire procedure, selecting new initial viewing angles, as many times as necessary until a total of 2000 satellite pairs have been generated.

\section*{An alternative estimate of the fraction of satellites in planes}

With the parameter selections detailed in the text, and setting $\Delta v_{\rm min} = \sqrt{2} \times 25\kms$, there are 380 galaxy systems in the SDSS. Using $\alpha=8^\circ$, we find $20/380$ pairs to have anti-correlated velocities, and $2/380$ to have correlated velocities, i.e. $22/380=5.8$\% of all pairs are found with this tolerance angle. With the unaltered Millennium II simulation (0\% in a disk) we find 4.7\% of pairs with $\alpha=8^\circ$; this fraction rises to 4.9\% with a 50\% disk component, and to 6.7\% with a 100\% disk component. This suggests that the fraction of satellites in a planar component within the SDSS is greater than 50\%, consistent with the estimate given in the text, but the simplicity of the disk model for the planar component prevents us from drawing strong conclusions from this comparison.

\section*{References}

\begin{thebibliography}{10}
\expandafter\ifx\csname url\endcsname\relax
  \def\url#1{\texttt{#1}}\fi
\expandafter\ifx\csname urlprefix\endcsname\relax\def\urlprefix{URL }\fi
\providecommand{\bibinfo}[2]{#2}
\providecommand{\eprint}[2][]{\url{#2}}

\bibitem{2005A&A...431..517K}
\bibinfo{author}{Kroupa, P.}, \bibinfo{author}{Theis, C.} \&
  \bibinfo{author}{Boily, C.~M.}
\newblock \bibinfo{title}{{The great disk of Milky-Way satellites and
  cosmological sub-structures}}.
\newblock \emph{\bibinfo{journal}{Astron. Astrophys.}} \textbf{\bibinfo{volume}{431}},
  \bibinfo{pages}{517--521} (\bibinfo{year}{2005}).

\bibitem{2012MNRAS.423.1109P}
\bibinfo{author}{Pawlowski, M.~S.}, \bibinfo{author}{Pflamm-Altenburg, J.} \&
  \bibinfo{author}{Kroupa, P.}
\newblock \bibinfo{title}{{The VPOS: a vast polar structure of satellite
  galaxies, globular clusters and streams around the Milky Way}}.
\newblock \emph{\bibinfo{journal}{Mon. Not. R. Astron. Soc.}} \textbf{\bibinfo{volume}{423}},
  \bibinfo{pages}{1109--1126} (\bibinfo{year}{2012}).

\bibitem{2013MNRAS.435.1928P}
\bibinfo{author}{Pawlowski, M.~S.}, \bibinfo{author}{Kroupa, P.} \&
  \bibinfo{author}{Jerjen, H.}
\newblock \bibinfo{title}{{Dwarf galaxy planes: the discovery of symmetric
  structures in the Local Group}}.
\newblock \emph{\bibinfo{journal}{Mon. Not. R. Astron. Soc.}} \textbf{\bibinfo{volume}{435}},
  \bibinfo{pages}{1928--1957} (\bibinfo{year}{2013}).

\bibitem{2013MNRAS.435.2116P}
\bibinfo{author}{Pawlowski, M.~S.} \& \bibinfo{author}{Kroupa, P.}
\newblock \bibinfo{title}{{The rotationally stabilized VPOS and predicted
  proper motions of the Milky Way satellite galaxies}}.
\newblock \emph{\bibinfo{journal}{Mon. Not. R. Astron. Soc.}} \textbf{\bibinfo{volume}{435}},
  \bibinfo{pages}{2116--2131} (\bibinfo{year}{2013}).

\bibitem{2013Natur.493...62I}
\bibinfo{author}{Ibata, R.~A.} \emph{et~al.}
\newblock \bibinfo{title}{{A vast, thin plane of corotating dwarf galaxies
  orbiting the Andromeda galaxy}}.
\newblock \emph{\bibinfo{journal}{Nature}} \textbf{\bibinfo{volume}{493}},
  \bibinfo{pages}{62--65} (\bibinfo{year}{2013}).

\bibitem{2013ApJ...766..120C}
\bibinfo{author}{Conn, A.~R.} \emph{et~al.}
\newblock \bibinfo{title}{{The Three-dimensional Structure of the M31 Satellite
  System; Strong Evidence for an Inhomogeneous Distribution of Satellites}}.
\newblock \emph{\bibinfo{journal}{Astrophys. J.}} \textbf{\bibinfo{volume}{766}},
  \bibinfo{pages}{120} (\bibinfo{year}{2013}).

\bibitem{2014ApJ...784L...6I}
\bibinfo{author}{Ibata, R.~A.} \emph{et~al.}
\newblock \bibinfo{title}{{A Thousand Shadows of Andromeda: Rotating Planes of
  Satellites in the Millennium-II Cosmological Simulation}}.
\newblock \emph{\bibinfo{journal}{Astrophys. J.}} \textbf{\bibinfo{volume}{784}},
  \bibinfo{pages}{L6} (\bibinfo{year}{2014}).

\bibitem{1976MNRAS.174..695L}
\bibinfo{author}{Lynden-Bell, D.}
\newblock \bibinfo{title}{{Dwarf galaxies and globular clusters in high
  velocity hydrogen streams}}.
\newblock \emph{\bibinfo{journal}{Mon. Not. R. Astron. Soc.}} \textbf{\bibinfo{volume}{174}},
  \bibinfo{pages}{695--710} (\bibinfo{year}{1976}).

\bibitem{AdelmanMcCarthy:2006iz}
\bibinfo{author}{Adelman-McCarthy, J.~K.} \emph{et~al.}
\newblock \bibinfo{title}{{The Fourth Data Release of the Sloan Digital Sky
  Survey}}.
\newblock \emph{\bibinfo{journal}{Astrophys. J. Supp.}} \textbf{\bibinfo{volume}{162}},
  \bibinfo{pages}{38} (\bibinfo{year}{2006}).

\bibitem{2007MNRAS.374.1125M}
\bibinfo{author}{Metz, M.}, \bibinfo{author}{Kroupa, P.} \&
  \bibinfo{author}{Jerjen, H.}
\newblock \bibinfo{title}{{The spatial distribution of the Milky Way and
  Andromeda satellite galaxies}}.
\newblock \emph{\bibinfo{journal}{Mon. Not. R. Astron. Soc.}} \textbf{\bibinfo{volume}{374}},
  \bibinfo{pages}{1125--1145} (\bibinfo{year}{2007}).

\bibitem{2008ApJ...680..287M}
\bibinfo{author}{Metz, M.}, \bibinfo{author}{Kroupa, P.} \&
  \bibinfo{author}{Libeskind, N.~I.}
\newblock \bibinfo{title}{{The Orbital Poles of Milky Way Satellite Galaxies: A
  Rotationally Supported Disk of Satellites}}.
\newblock \emph{\bibinfo{journal}{Astrophys. J.}} \textbf{\bibinfo{volume}{680}},
  \bibinfo{pages}{287--294} (\bibinfo{year}{2008}).

\bibitem{2009MNRAS.394.2223M}
\bibinfo{author}{Metz, M.}, \bibinfo{author}{Kroupa, P.} \&
  \bibinfo{author}{Jerjen, H.}
\newblock \bibinfo{title}{{Discs of satellites: the new dwarf spheroidals}}.
\newblock \emph{\bibinfo{journal}{Mon. Not. R. Astron. Soc.}} \textbf{\bibinfo{volume}{394}},
  \bibinfo{pages}{2223--2228} (\bibinfo{year}{2009}).

\bibitem{2009Natur.461...66M}
\bibinfo{author}{McConnachie, A.~W.} \emph{et~al.}
\newblock \bibinfo{title}{{The remnants of galaxy formation from a panoramic
  survey of the region around M31}}.
\newblock \emph{\bibinfo{journal}{Nature}} \textbf{\bibinfo{volume}{461}},
  \bibinfo{pages}{66--69} (\bibinfo{year}{2009}).

\bibitem{2014ApJ...780..128I}
\bibinfo{author}{Ibata, R.~A.} \emph{et~al.}
\newblock \bibinfo{title}{{The Large-scale Structure of the Halo of the
  Andromeda Galaxy. I. Global Stellar Density, Morphology and Metallicity
  Properties}}.
\newblock \emph{\bibinfo{journal}{Astrophys. J.}} \textbf{\bibinfo{volume}{780}},
  \bibinfo{pages}{128} (\bibinfo{year}{2014}).

\bibitem{2012ApJ...758...11C}
\bibinfo{author}{Conn, A.~R.} \emph{et~al.}
\newblock \bibinfo{title}{{A Bayesian Approach to Locating the Red Giant Branch
  Tip Magnitude. II. Distances to the Satellites of M31}}.
\newblock \emph{\bibinfo{journal}{Astrophys. J.}} \textbf{\bibinfo{volume}{758}},
  \bibinfo{pages}{11} (\bibinfo{year}{2012}).

\bibitem{2012ApJ...752...45T}
\bibinfo{author}{Tollerud, E.~J.} \emph{et~al.}
\newblock \bibinfo{title}{{The SPLASH Survey: Spectroscopy of 15 M31 Dwarf
  Spheroidal Satellite Galaxies}}.
\newblock \emph{\bibinfo{journal}{Astrophys. J.}} \textbf{\bibinfo{volume}{752}},
  \bibinfo{pages}{45} (\bibinfo{year}{2012}).

\bibitem{2013ApJ...768..172C}
\bibinfo{author}{Collins, M. L.~M.} \emph{et~al.}
\newblock \bibinfo{title}{{A Kinematic Study of the Andromeda Dwarf Spheroidal
  System}}.
\newblock \emph{\bibinfo{journal}{Astrophys. J.}} \textbf{\bibinfo{volume}{768}},
  \bibinfo{pages}{172} (\bibinfo{year}{2013}).

\bibitem{2013AJ....146..126C}
\bibinfo{author}{Chiboucas, K.}, \bibinfo{author}{Jacobs, B.~A.},
  \bibinfo{author}{Tully, R.~B.} \& \bibinfo{author}{Karachentsev, I.~D.}
\newblock \bibinfo{title}{{Confirmation of Faint Dwarf Galaxies in the M81
  Group}}.
\newblock \emph{\bibinfo{journal}{Astron. J.}} \textbf{\bibinfo{volume}{146}},
  \bibinfo{pages}{126} (\bibinfo{year}{2013}).

\bibitem{2013A&A...559L..11B}
\bibinfo{author}{Bellazzini, M.}, \bibinfo{author}{Oosterloo, T.},
  \bibinfo{author}{Fraternali, F.} \& \bibinfo{author}{Beccari, G.}
\newblock \bibinfo{title}{{Dwarfs walking in a row. The filamentary nature of
  the NGC 3109 association}}.
\newblock \emph{\bibinfo{journal}{Astron. Astrophys.}} \textbf{\bibinfo{volume}{559}},
  \bibinfo{pages}{L11} (\bibinfo{year}{2013}).

\bibitem{2013MNRAS.431.3543H}
\bibinfo{author}{Hammer, F.} \emph{et~al.}
\newblock \bibinfo{title}{{The vast thin plane of M31 corotating dwarfs: an
  additional fossil signature of the M31 merger and of its considerable impact
  in the whole Local Group}}.
\newblock \emph{\bibinfo{journal}{Mon. Not. R. Astron. Soc.}} \textbf{\bibinfo{volume}{431}},
  \bibinfo{pages}{3543--3549} (\bibinfo{year}{2013}).

\bibitem{2013pss5.book.1039W}
\bibinfo{author}{Walker, M.}
\newblock \bibinfo{title}{{Dark Matter in the Galactic Dwarf Spheroidal Satellites}}.
\newblock \emph{\bibinfo{journal}{in Planets, Stars and Stellar Systems, Vol. 5, Oswalt, T., Gilmore, G. Eds., Springer, Dordrecht}} \bibinfo{pages}{1039}
  (\bibinfo{year}{2013}).

\bibitem{2013MNRAS.436.2096S}
\bibinfo{author}{Shaya, E.}, \bibinfo{author}{Tully, B.}
\newblock \bibinfo{title}{{The formation of Local Group planes of galaxies}}.
\newblock \emph{\bibinfo{journal}{Mon. Not. R. Astron. Soc.}} \textbf{\bibinfo{volume}{436}},
  \bibinfo{pages}{2096--2119} (\bibinfo{year}{2013}).

\bibitem{2009MNRAS.398.1150B}
\bibinfo{author}{Boylan-Kolchin, M.}, \bibinfo{author}{Springel, V.},
  \bibinfo{author}{White, S. D.~M.}, \bibinfo{author}{Jenkins, A.} \&
  \bibinfo{author}{Lemson, G.}
\newblock \bibinfo{title}{{Resolving cosmic structure formation with the
  Millennium-II Simulation}}.
\newblock \emph{\bibinfo{journal}{Mon. Not. R. Astron. Soc.}} \textbf{\bibinfo{volume}{398}},
  \bibinfo{pages}{1150--1164} (\bibinfo{year}{2009}).

\bibitem{2013MNRAS.428.1351G}
\bibinfo{author}{Guo, Q.} \emph{et~al.}
\newblock \bibinfo{title}{{Galaxy formation in WMAP1 and WMAP7 cosmologies}}.
\newblock \emph{\bibinfo{journal}{Mon. Not. R. Astron. Soc.}} \textbf{\bibinfo{volume}{428}},
  \bibinfo{pages}{1351--1365} (\bibinfo{year}{2013}).

\bibitem{2011ApJS..192...18K}
\bibinfo{author}{Komatsu, E.} \emph{et~al.}
\newblock \bibinfo{title}{{Seven-year Wilkinson Microwave Anisotropy Probe
  (WMAP) Observations: Cosmological Interpretation}}.
\newblock \emph{\bibinfo{journal}{Astrophys. J. Supp.}} \textbf{\bibinfo{volume}{192}},
  \bibinfo{pages}{18} (\bibinfo{year}{2011}).

\bibitem{2014MNRAS.438.2916B}
\bibinfo{author}{Bahl, H.} \& \bibinfo{author}{Baumgardt, H.}
\newblock \bibinfo{title}{{A comparison of the distribution of satellite
  galaxies around Andromeda and the results of $\Lambda$CDM simulations}}.
\newblock \emph{\bibinfo{journal}{Mon. Not. R. Astron. Soc.}} \textbf{\bibinfo{volume}{438}},
  \bibinfo{pages}{2916--2923} (\bibinfo{year}{2014}).

\bibitem{2005AJ....129.2562B}
\bibinfo{author}{Blanton, M.~R.} \emph{et~al.}
\newblock \bibinfo{title}{{New York University Value-Added Galaxy Catalog: A
  Galaxy Catalog Based on New Public Surveys}}.
\newblock \emph{\bibinfo{journal}{Astron. J.}} \textbf{\bibinfo{volume}{129}},
  \bibinfo{pages}{2562--2578} (\bibinfo{year}{2005}).

\bibitem{2013arXiv1303.5076P}
\bibinfo{author}{Ade, P.A.R.} \emph{et~al.}
\newblock \bibinfo{title}{{Planck 2013 results. XVI. Cosmological parameters}}.
\newblock \emph{\bibinfo{journal}{arXiv}} \bibinfo{pages}{5076}
  (\bibinfo{year}{2013}).
\newblock \eprint{1303.5076v2}.

\bibitem{2002AJ....124.1810S}
\bibinfo{author}{Strauss, M.~A.} \emph{et~al.}
\newblock \bibinfo{title}{{Spectroscopic Target Selection in the Sloan Digital
  Sky Survey: The Main Galaxy Sample}}.
\newblock \emph{\bibinfo{journal}{Astron. J.}} \textbf{\bibinfo{volume}{124}},
  \bibinfo{pages}{1810--1824} (\bibinfo{year}{2002}).

\bibitem{2011MNRAS.414.2446M}
\bibinfo{author}{McMillan, P.~J.}
\newblock \bibinfo{title}{{Mass models of the Milky Way}}.
\newblock \emph{\bibinfo{journal}{Mon. Not. R. Astron. Soc.}} \textbf{\bibinfo{volume}{414}},
  \bibinfo{pages}{2446--2457} (\bibinfo{year}{2011}).

\end{thebibliography}

\begin{thebibliography}{1}
\expandafter\ifx\csname url\endcsname\relax
  \def\url#1{\texttt{#1}}\fi
\expandafter\ifx\csname urlprefix\endcsname\relax\def\urlprefix{URL }\fi
\providecommand{\bibinfo}[2]{#2}
\providecommand{\eprint}[2][]{\url{#2}}

\bibitem[31]{Navarro:1997if}
\bibinfo{author}{Navarro, J.~F., Frenk, C.~S., and White, S. D.~M.}
\newblock \bibinfo{title}{{A Universal Density Profile from Hierarchical Clustering}}.
\newblock \emph{\bibinfo{journal}{Astrophys. J.}} \textbf{\bibinfo{volume}{490}},
  \bibinfo{pages}{493--508} (\bibinfo{year}{1997}).


\bibitem[32]{2008MNRAS.384.1459L}
\bibinfo{author}{Li, Y.-S. and White, S. D.~M.}
\newblock \bibinfo{title}{{Masses for the Local Group and the Milky Way}}.
\newblock \emph{\bibinfo{journal}{Mon. Not. R. Astron. Soc.}} \textbf{\bibinfo{volume}{384}},
  \bibinfo{pages}{1459--1468} (\bibinfo{year}{2008}).



\end{thebibliography}

\clearpage

\begin{figure}
\includegraphics[angle=0, bb=0 0 550 483, clip,  width=16cm]{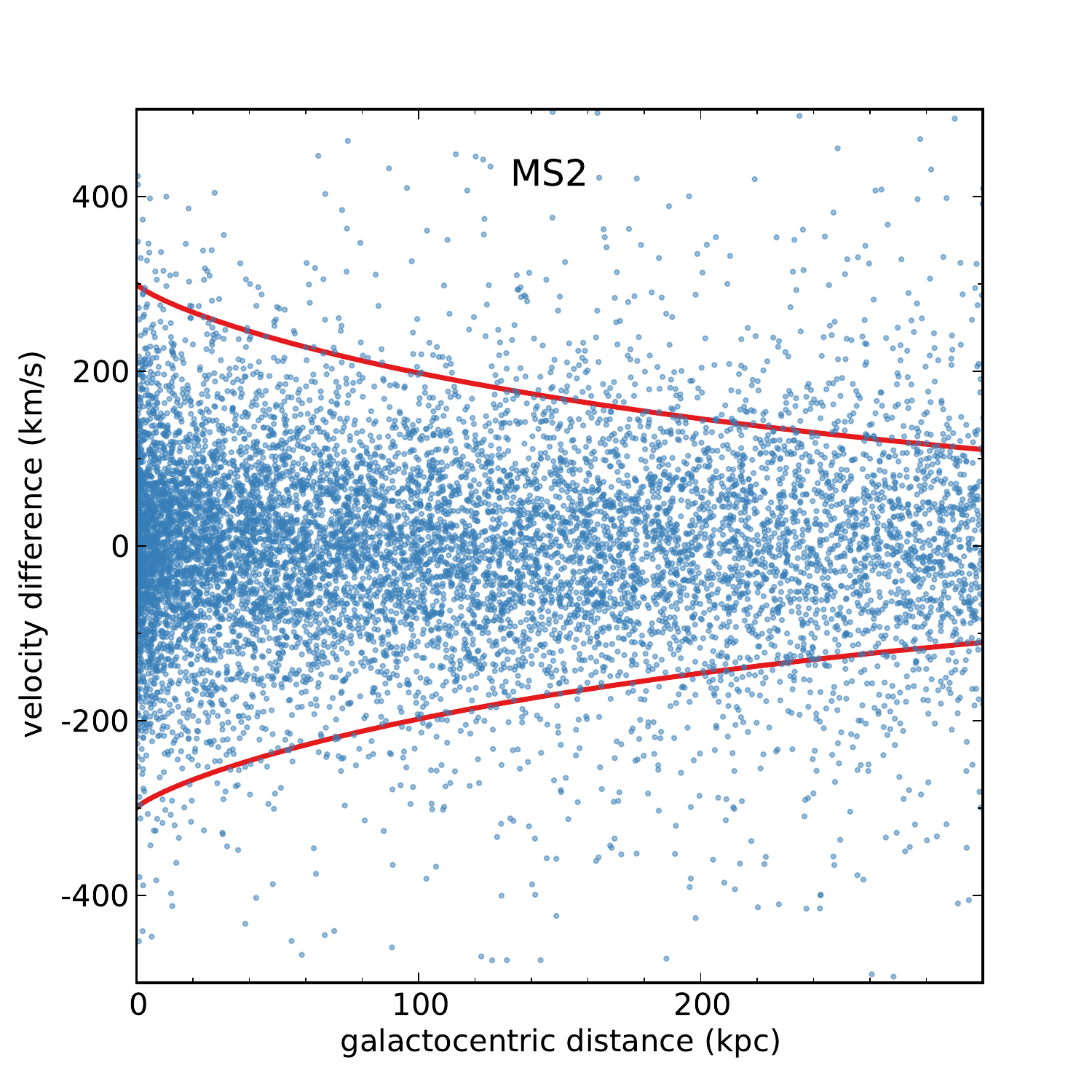}
{{\bf Extended Data Figure 1} Adopted velocity envelope relation. Dots mark the distance-velocity distribution of satellites in the MS2 simulation that surround isolated host galaxies of similar luminosity and mass to the Milky-Way\cite{2008MNRAS.384.1459L}. The empirical envelope relation shown in red ($300 \exp[-(300\kpc/R)^{0.8}]\kms$) is used in our analysis as a means to reduce contamination from velocity outliers.}
\end{figure}

\end{document}